\documentclass[prl,twocolumn,groupedaddress,superscriptaddress]{revtex4-1}
\usepackage{eurosym}
\usepackage{amsfonts}
\usepackage{amsmath}
\usepackage{amssymb}
\usepackage{graphicx}
\usepackage{float}

\begin{document}

\title{Opening a nodal gap by fluctuating spin-density-wave in lightly doped La$_{2-x}$Sr$_x$CuO$_4$}

\author{Itzik Kapon}
\affiliation{Department of Physics, Technion - Israel Institute of Technology, Haifa, 3200003, Israel}

\author{David S. Ellis}
\affiliation{Department of Physics, Technion - Israel Institute of Technology, Haifa, 3200003, Israel}

\author{Gil Drachuck}
\affiliation{Department of Physics, Technion - Israel Institute of Technology, Haifa, 3200003, Israel}
\affiliation{Department of Physics and Astronomy, Iowa State University, Ames, IA 50011, USA and Ames Laboratory, Iowa State University, Ames, IA 50011, USA}

\author{Galina Bazalitski}
\affiliation{Department of Physics, Technion - Israel Institute of Technology, Haifa, 3200003, Israel}

\author{Eugen Weschke}
\affiliation{Helmholtz-Zentrum Berlin f\"{u}r Materialien und Energie, Albert-Einstein-Stra\ss e 15, D-12489 Berlin, Germany}

\author{Enrico Schierle}
\affiliation{Helmholtz-Zentrum Berlin f\"{u}r Materialien und Energie, Albert-Einstein-Stra\ss e 15, D-12489 Berlin, Germany}

\author{J\"{o}rg Strempfer}
\affiliation{Deutsches  Elektronen-Synchrotron  DESY,  Notkestra\ss e  85,  22607  Hamburg,  Germany}

\author{Christof Niedermayer}
\affiliation{Laboratory for Neutron Scattering, Paul Scherrer Institute, CH-5232 Villigen PSI, Switzerland}

\author{Amit Keren}
\affiliation{Department of Physics, Technion - Israel Institute of Technology, Haifa, 3200003, Israel}

\date{\today }

\begin{abstract}

We investigate whether the spin or charge degrees of freedom are responsible for the nodal gap in underdoped cuprates by performing inelastic neutron scattering and x-ray diffraction measurements on La$_{2-x}$Sr$_x$CuO$_4$, which is on the edge of the antiferromagnetic phase. We found that fluctuating incommensurate spin-density-wave (SDW) with a the bottom part of an hourglass dispersion exists even in this magnetic sample. The strongest component of these fluctuations diminishes at the same temperature where the nodal gap opens. X-ray scattering measurements on the same crystal show no signature of charge-density-wave (CDW). Therefore, we suggest that the nodal gap in the electronic band of this cuprate opens due to fluctuating SDW with no contribution from CDW.

\end{abstract}

\maketitle



There are several indications by now from angle-resolved photoemission spectroscopy (ARPES) that in underdoped cuprates a gap opens at the Fermi surface in the diagonal (nodal) direction~\cite{Drachuck,AHarter,vishik2012,razzoli2013evolution,peng2013disappearance}. In La$_{2-x}$Sr$_x$CuO$_4$ (LSCO) this nodal gap (NG) extends to $x=8\%$. At doping around $x=12.5\%$ samples develop a charge-density-wave (CDW) below $T\approx100$~K \cite{croft14}. Traces of antiferromagnetism (AFM) in the form of spin-density-waves (SDW) \cite{matsuda2002} or spin-glass \cite{Niedermayer98} appear at doping up to $x=12.5\%$ and temperatures $T\approx10$~K. It is therefore natural to speculate that one of these symmetry breaking phenomena is responsible for the opening of a nodal gap. In this work, we would like to clarify which one is the most likely. Our strategy is to carefully examine a sample which is known to at least have both AFM and SDW order, and opens a nodal gap at low temperatures. The sample is LSCO with $x=1.92\%$~\cite{Drachuck}.

Previous neutron diffraction measurements on LSCO $x=1.92\%$ \cite{Drachuck} showed a magnetic Bragg peak at the AFM wave vector $\textbf{Q}_{AF}$ below $T=140$~K, and two satellites that stand for static SDW order (on top of the AFM one). The satellites appear below $T=30$~K. Like in Matsuda et al. ~\cite{matsuda2002}, there are two domains in the sample. We focus on one of them, in which the AFM peak is observed when scanning near (1,0,0), with no contribution from SDW. In contrast, the SDW peaks are observed when scanning near (0,1,0), with no contribution from the AFM peak. Neutron scattering detects the component of spin fluctuations perpendicular to the momentum transfer $\textbf{q}$ ~\cite{squires2012introduction}. Hence, the SDW fluctuations are perpendicular to the AFM order. ARPES measurements on the same sample found that a nodal gap opens below $T_{NG}=45$~K~\cite{Drachuck}. Even though there is a temperature mismatch between the NG and SDW appearance, the two phenomena might be related. Moreover, CDW in LSCO $x=1.92\%$ is expected to be very weak \cite{capati2015}, and indeed this sample is out of the CDW dome ~\cite{croft14,Hucker14}. Therefore, \textit{a priori}, CDW is not expected to generate the nodal gap.

Here we add to the available ARPES and neutron diffraction data, inelastic neutron scattering (INS) and x-ray diffraction data on the same piece of LSCO $x=1.92\%$. We show that the fluctuating SDW amplitude of the frequency where it is the strongest, decreases at a temperature equal to $T_{NG}$ within experimental error. In addition, we could not find any indications for CDW in our sample. We argue that these findings explain the previously measured $15$~K discrepancy between the SDW freezing and the opening of a NG, and tie the latter to fluctuating SDW.

\begin{figure*}[h!t]
	\begin{center}
		\includegraphics[trim=0cm 0cm 0.5cm 0cm ,clip=true,width=18cm]{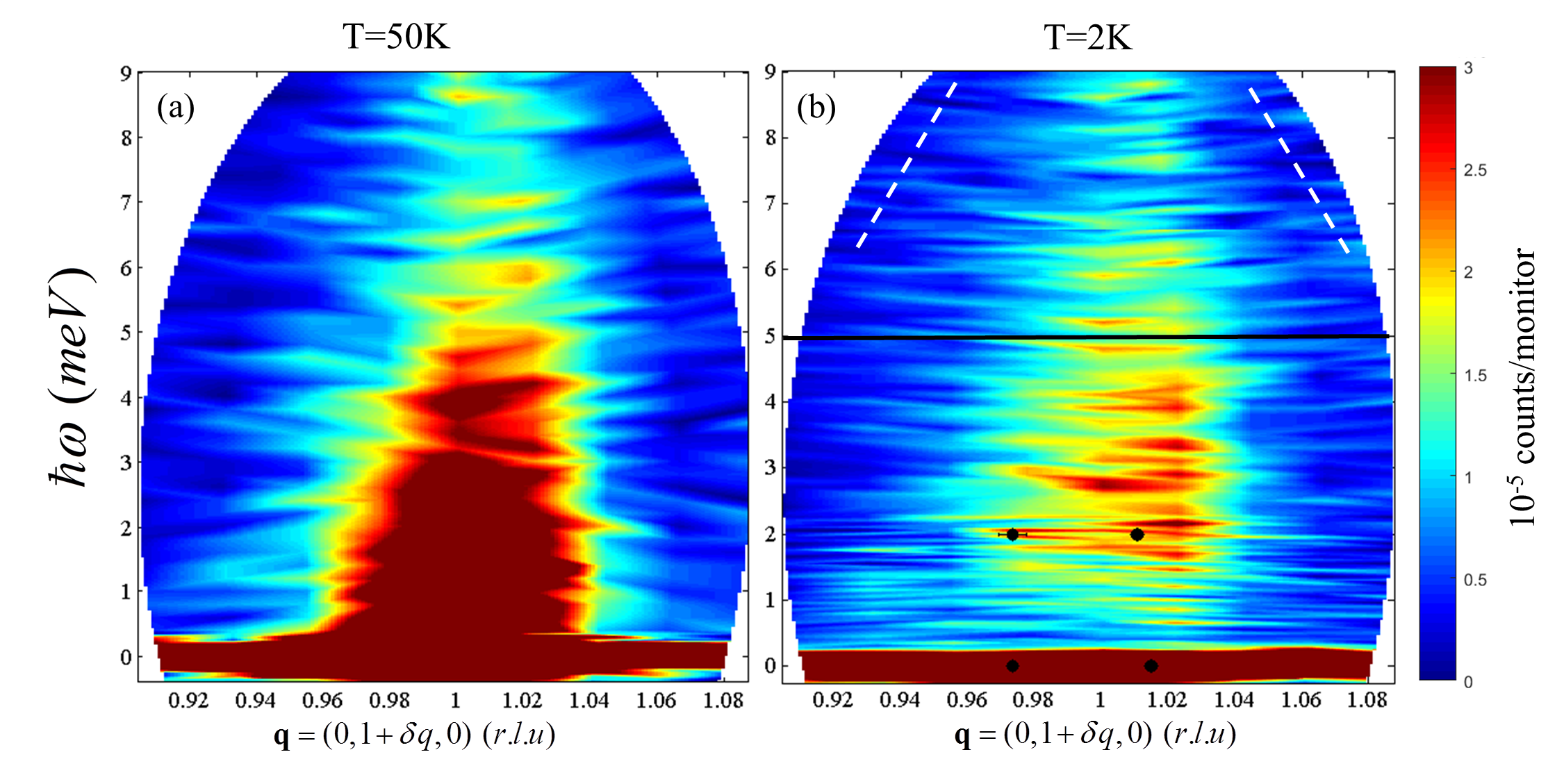}
	\end{center}
	\caption{\textbf{SDW dispersion.}
		False color map of normalized intensity as a function of neutron energy transfer $\hbar \omega$ and momentum transfer $\textbf{q}$ at $T=50$~K (a) and $T=2$~K (b). The raw data is interpolated. The black horizontal line in panel (b) demonstrates a constant energy cut along which the intensity is integrated and plotted in Fig.~\ref{fig:edep}. Dashed white lines in panel (b) represent cuts along which the background is determined. The black symbols indicate the center of the $\hbar \omega =0$ and $2$~meV peaks demonstrated in Fig.~\ref{fig:qdep}, and make the bottom part of an hourglass.}
	\label{fig:dispersion}
\end{figure*}

\begin{figure}[tbph]
		\includegraphics[trim=1cm 2cm 1cm 2.5cm ,clip=true,width=\columnwidth]{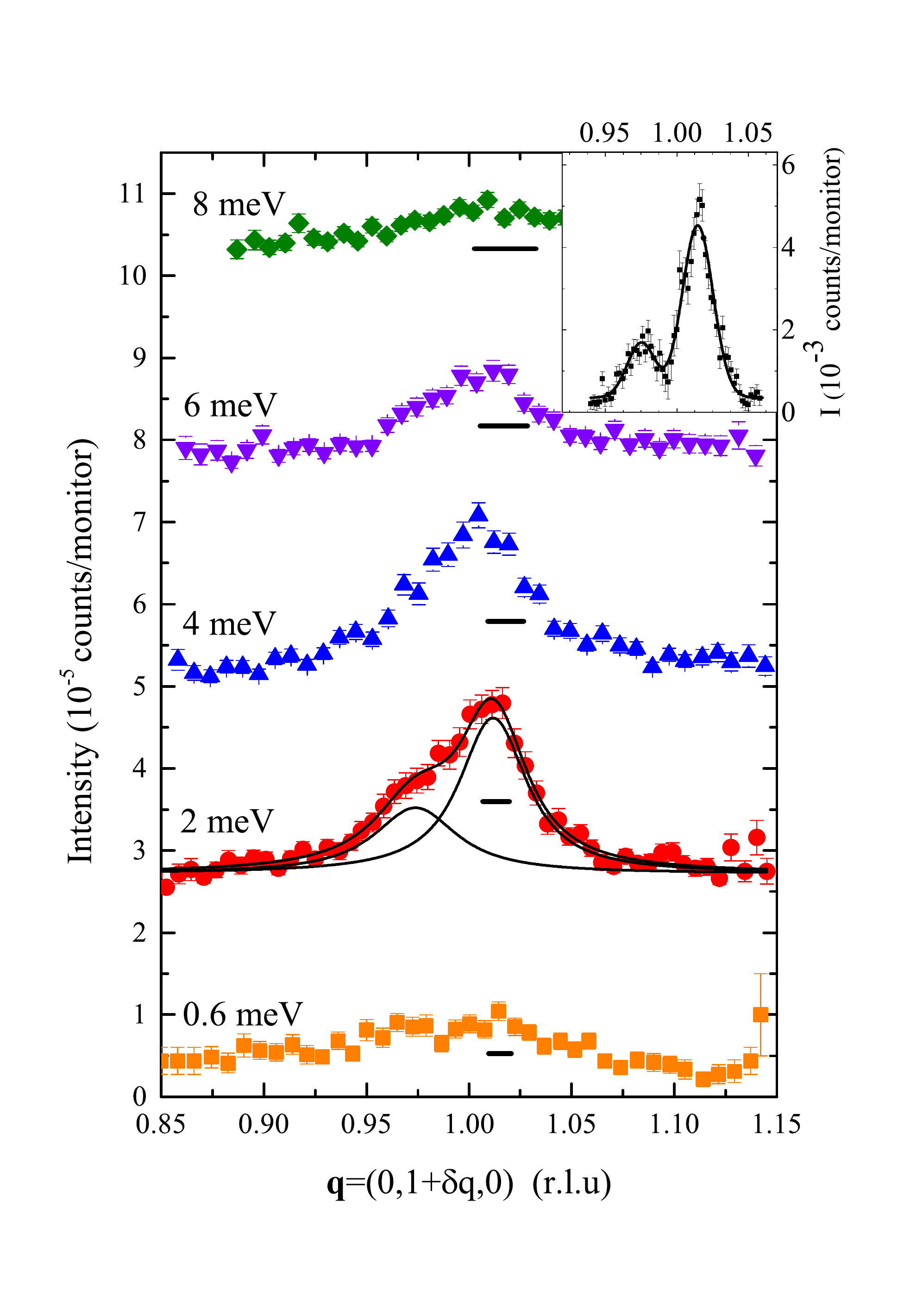}
	\caption{\textbf{Evolution of the SDW peaks with energy at $T=2$~K.}
		Momentum scan along k centered at (0,1,0) for different energy transfers at $T=2$~K. Scans are shifted consecutively by $2.5\times10^{-5}$ counts/monitor for clarity. Inset: SDW elastic peaks for the same $\textbf{q}$ scan also at $T=2$~K. Background from higher temperature was subtracted. For energies of $\hbar \omega=0$ and $2$~meV, a sum of two Voigt functions is fitted to the data (solid black lines). The peak separation for $\hbar \omega=2$~meV is 0.04 r.l.u, as in the $\hbar \omega=0$ case (see inset). Black horizontal lines represents the instrumental resolution.   
	}
	\label{fig:qdep}
\end{figure}

The neutron experiment was performed at Rita-II, the cold neutrons triple axis spectrometer at the Paul Scherrer Institut. Throughout this paper, we work in orthorhombic notation, with cell parameters $a=5.344$~\AA, $b=5.421$~\AA\  and $c=13.14$~\AA\  at $T=2$~K. In this notation, the tetragonal 2D $\textbf{Q}_{AF}=\left(1/2,1/2,0\right)$ is equivalent to $\left( 0,1,0\right)$ in reciprocal lattice units (r.l.u) of $2\pi/a$. More information is available in the Methods section. In Fig.~\ref{fig:dispersion} we present a false color map of neutron counts versus energy transfer $\hbar \omega$ and momentum transfer $\textbf{q}$. The raw data, in this figure alone, is interpolated for presentation purpose. Data is presented at two temperatures, $2$~K and $50$~K, which are below and above the freezing temperature of the incommensurate magnetic order of $30$~K ~\cite{Drachuck}. In both cases, strong intensity is observed at $\hbar \omega=0$. This is due to high order contamination of the incoming beam scattering from a nuclear Bragg peak at $(0,2,0)$, despite the use of Br filter. Around $(0,1,0)$, the intensity extends to energy transfers as high as 8 meV for both temperatures, in a cone shape, which is in fact a poorly-resolved bottom part of an hourglass. This will be demonstrated subsequently. The scattering intensity is stronger at elevated temperatures. Interestingly, at $T=2$~K spectral weight is missing at low energies, suggesting the presence of a soft gap for spin excitations. A similar spectrum, including the gap, was observed at the fully developed hourglass dispersion of La$_{1.875}$Ba$_{0.125}$CuO$_4$~\cite{tranquada2004}, La$_{1.88}$Sr$_{0.12}$CuO$_4$ ~\cite{matsuda2008,Romer13}, and La$_{1.6}$Sr$_{0.4}$CoO$_4$ ~\cite{drees2013hour}. 

q-scans at specific constant energies at $T=2$~K are presented in Fig.~\ref{fig:qdep}, showing the evolution of the SDW peaks with energy transfer. The intensities are shifted vertically for clarity. At $\hbar \omega=0.6$~meV, some intensity is detected around $(0,1,0)$ above the background. However, this could stem from the tail of the high order contamination. At $\hbar \omega=2$~meV two clear peaks appear 

For fitting, the instrument was modeled using Popovici ResCal5 ~\cite{popovici1975resolution}, and the resolution was calculated. Black horizontal lines in Fig.~\ref{fig:qdep} represent the q-resolution at each energy. This was taken into account as a constant width Gaussian at each energy, which was convoluted with a Lorentzian (Voigt function). The fit with two Voigt functions is demonstrated in Fig.~\ref{fig:qdep} by solid lines. The fit to the $\hbar \omega=2$~meV data indicates a peak separation of 0.04 r.l.u. The same separation is found in the elastic peaks ~\cite{Drachuck}, as demonstrated in the inset. The peaks centers are illustrated in Fig.~\ref{fig:dispersion}(b) by the solid points. The static and dynamic SDW correlation lengths, determined from the peaks width, are $85\pm12~\AA$ and $44\pm5~\AA$ respectively. With increasing energy to $4$~meV and then to $6$~meV, the two peaks are no longer resolved. However, the measured peak is asymmetric because of the two underlying incommensurate peaks coming closer together. At $8$~meV the intensity diminishes. This behavior reminds two ``legs" dispersing downwards from some crossing energy as in the hourglass.

\begin{figure}[tbph]
		\includegraphics[trim=1cm 1cm 1cm 1.2cm ,clip=true,width=\columnwidth]{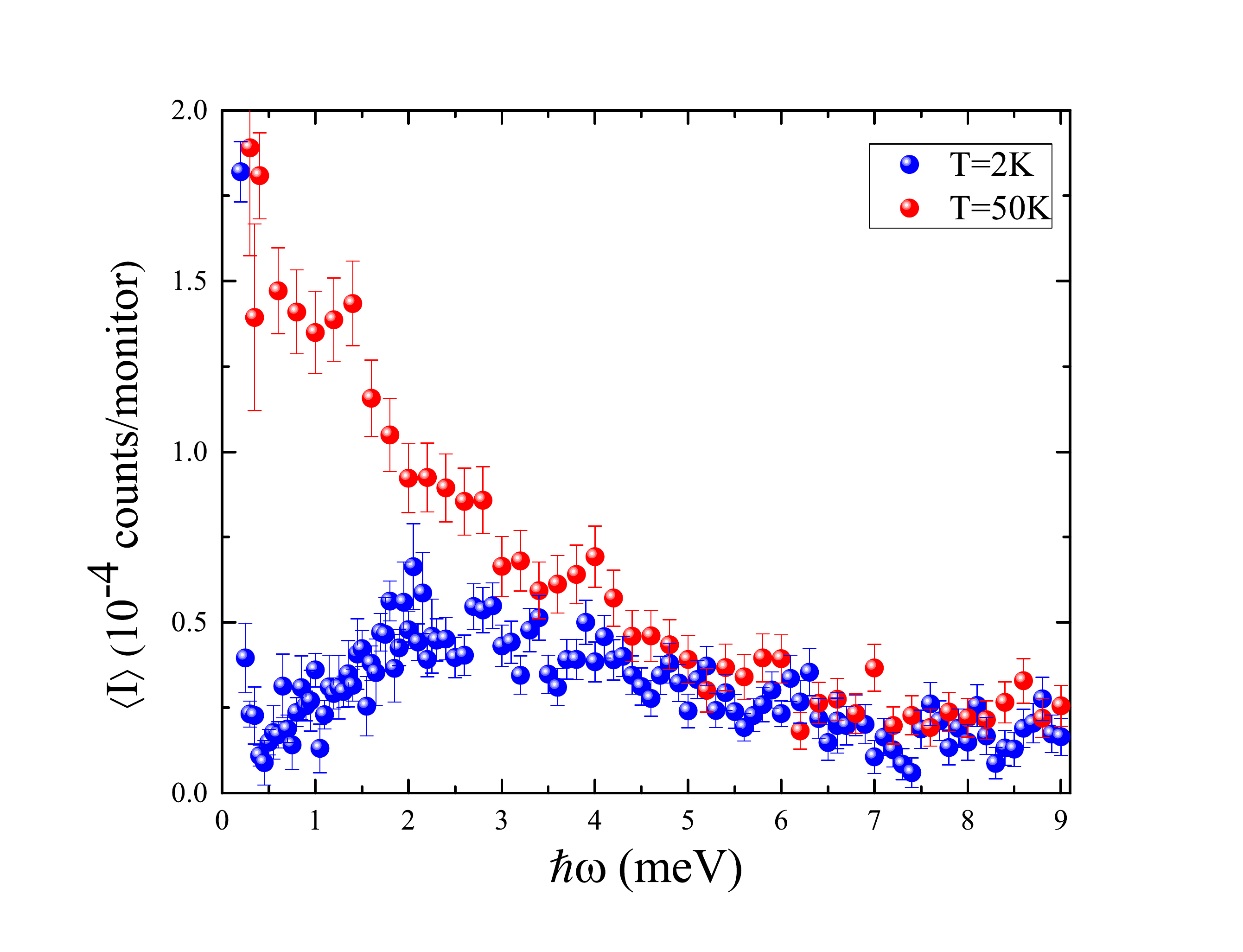}
	\caption{\textbf{q-integrated intensity vs. neutron energy transfer at low (2 K) and high (50 K) temperatures.}
		Integrated intensity is calculated for each energy as sum of the counts over $\textbf{q}$ along horizontal lines like the one shown in Fig.~\ref{fig:dispersion}. Background is estimated from the counts along the two dashed lines shown in Fig.~\ref{fig:dispersion}(a) and subtracted from the raw data.
	}
	\label{fig:edep}
\end{figure}

To further investigate the inelastic behavior, we sum the intensity over $\textbf{q}$ at constant energy cuts. The horizontal line in Fig.~\ref{fig:dispersion}(b) presents one such cut. Background contribution is estimated from the data along the dashed diagonal lines in Fig.~\ref{fig:dispersion}(b), and subtracted. Fig.~\ref{fig:edep} presents the background subtracted \textbf{q}-integrated intensity versus energy transfer $ \langle I \rangle (\omega)=\sum_\textbf{q} I(\textbf{q},\omega)$, starting from $\hbar\omega = 0.15$ meV to avoid the high intensity elastic peak. At $T=50$~K, $\langle I \rangle (\omega)$ monotonically grows as the frequency decreases. In contrast, at $T=2$~K, $\langle I \rangle (\omega)$ reaches a maximum at some $\hbar\omega_{max}$ between 2 and 3~meV, and drops towards $\hbar\omega=0$, although residual elastic scattering intensity is observed near $\hbar\omega=0$. Measurements on La$_{2-x}$Ba$_x$CuO$_4$ with $0.0125\le x \le 0.035$  which were limited to energies below 1meV agree with our results ~\cite{Wagman2013}. This plot demonstrates more clearly the aforementioned soft gap in spin excitations which develops at low temperatures.

\begin{figure}[tbph]
	\begin{center}
		\includegraphics[trim=0cm 3cm 0cm 1cm ,clip=true,width=\columnwidth]{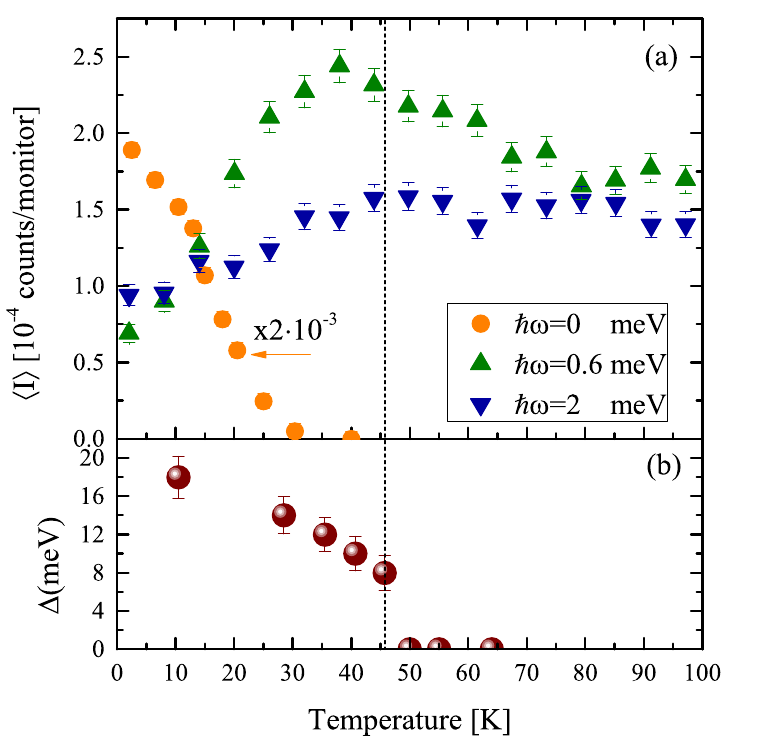}
	\end{center}
	\caption{\textbf{Temperature dependence of all experimental parameters.}
		(a) Elastic and inelastic incommensurate SDW intensities at different energies from neutron scattering. (b) ARPES measurement of the nodal gap at $k_F$ \cite{Drachuck}. The dashed vertical line emphasizes the fact that the nodal gap opens when the amplitude of dynamic spin fluctuations at $\hbar\omega\approx~2$meV decreases.
	}
	\label{fig:tdep}
\end{figure}

We summarize the available data on LSCO $x=1.92\%$ in Fig.~\ref{fig:tdep}(a). In this figure we show the temperature dependence of the $\textbf{q}$-integrated scattering intensity at three different energies. The data at $\hbar \omega=0$ is taken from Ref.~\cite{Drachuck} and multiplied by $2\times 10^{-3}$ for clarity. It shows that a long range static SDW appears at a temperature of $30$~K. The intensity at $\hbar \omega=0.6$~meV increases as the temperature is lowered, peaks at $38$~K, and then decreases. This result demonstrates that dynamically fluctuating SDW at $\hbar\omega >0$ diminishes upon cooling before long range static incommensurate order develops. The same effect, although less sharp, is observed for $\hbar \omega=2$~meV at $45$~K. 

Figure ~\ref{fig:tdep}(b) depicts the temperature dependence of the nodal gap from Ref.~\cite{Drachuck} as measured by ARPES. This gap opens at $T_{NG}=45$~K, which is the same temperature where the spectral density at $\hbar\omega_{max}$ begins to diminish. The maximum electronic gap value $\Delta$ agrees with isolated dopant-hole bound state calculations~\cite{sushkov2005}. We note that $\hbar\omega_{max}$ and $k_B T_{NG}$ are of the same order of magnitude. Our result indicates a strong link between the dynamically fluctuating SDW and the nodal gap.

In order to investigate whether CDW plays a role in the nodal gap \cite{Berg08}, we conducted a search for CDW in this sample by two different methods: off resonance x-ray diffraction (XRD) and resonance elastic x-ray scattering (REXS). The experiments were done at PETRA III on the P09 beam-line and at BESSY on the UE46-PGM1 beam-line, respectively. In REXS, the background subtraction is not trivial, so we only present here our XRD data. Nonetheless, the final conclusion from both methods is the same.

\begin{figure}[tbph]
	\begin{center}
		\includegraphics[trim=1.5cm 1cm 3cm 0cm ,clip=true,width=\columnwidth]{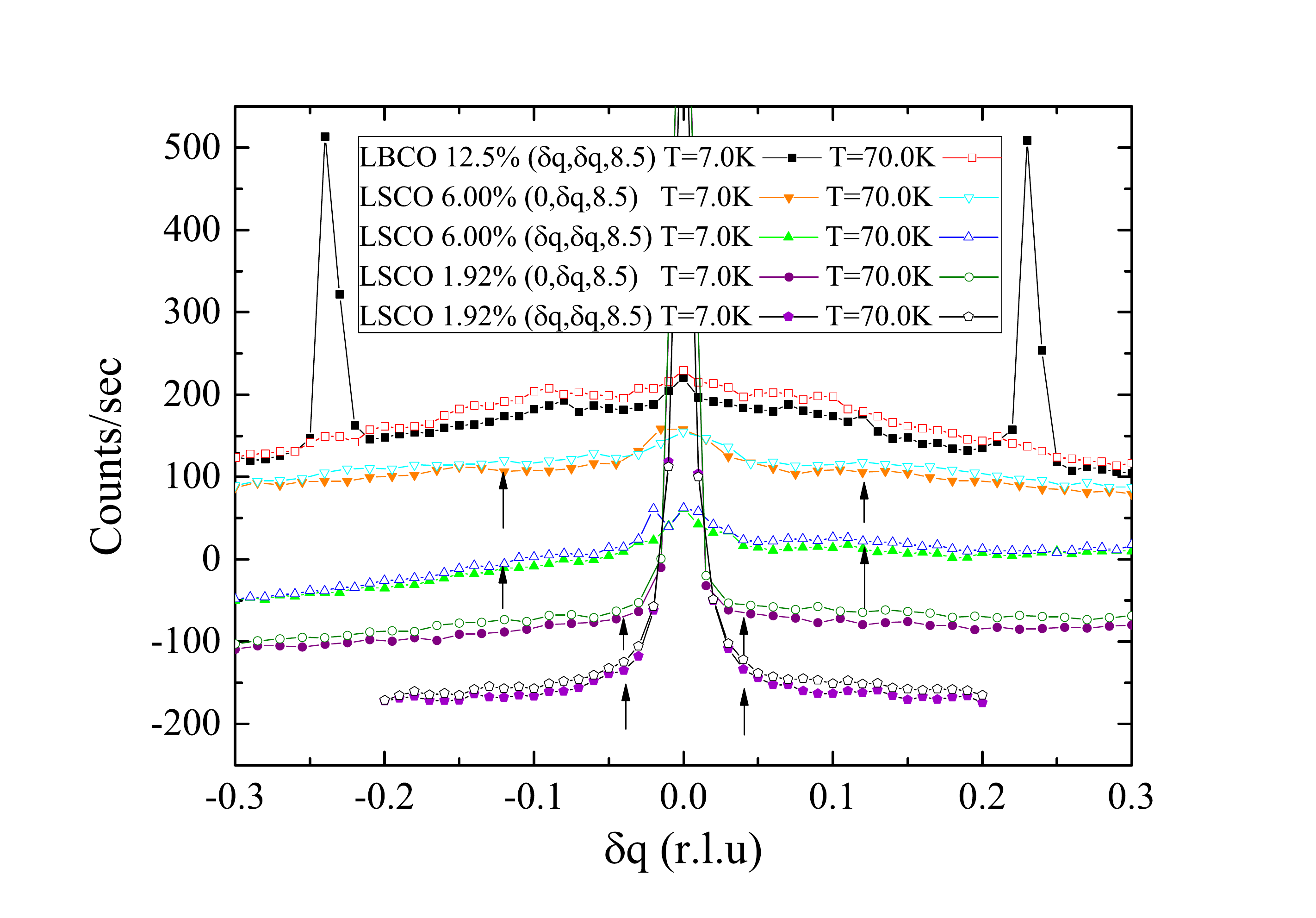}
	\end{center}
	\caption{\textbf{Hard x-ray diffraction on three different samples: }LSCO with $x=1.92\%$ and $x=6\%$, and LBCO $x=12.5\%$. Scans are done in two different orientations and two different temperatures. CDW is detected only in LBCO. }
	\label{fig:cdw}
\end{figure}

In Fig.~\ref{fig:cdw} we show results from LSCO samples with $x=1.92 \%$, $x=6.0\%$, and La$_{2-x}$Ba$_x$CuO$_4$ (LBCO) $x=12.5\%$. The data sets are shifted vertically for clarity. The LBCO sample is used as a test case, since it has well established CDW and presents strong diffraction peaks. The measurements were taken at $7$~K and at $70$~K, which are below and above the CDW critical temperature of LBCO~\cite{Tranquada2008cdw}. We performed two types of scans: a ``stripes" scan along $(0,\delta q,8.5)$ direction and a ``checkerboard" scan along $(\delta q,\delta q,8.5)$ direction. We chose to work at $l=8.5$ to minimize contribution from a Bragg peak at $l=8$ or $l=9$. For LBCO at $T=7$~K, there is a clear CDW peak at $\delta q=\pm 0.24$ in the ``checkerboard" scan, which is absent at high temperatures. In contrast, for the LSCO samples there is no difference between the signal at high and low temperatures. Since $\delta q$ of the CDW peak depends on doping, in our sample it is expected to be close to $\delta q=0$, where a tail of the Bragg peak could potentially obscure the CDW peak. Arrows in Fig.~\ref{fig:cdw} show where we might expect the CDW peaks, should they appear, based on linear scaling with doping. These positions are out of the $\delta q=0$ peak tail, and not obscured. Thus, although we are in experimental conditions appropriate to find a CDW, it is not observed within our sensitivity. In fact, CDW is even absent at higher doping as demonstrated by our experiment with LSCO $x=6\%$ sample. We observed the same null-result with the REXS experiment. It is important to mention that hourglass excitations with no stripe-like CDW were observed previously \cite{drees2013hour}.

Our main results are as follows: we find the bottom part of an hourglass dispersion inside the AFM phase of LSCO. The hourglass does not start from zero energy, but has a soft gap from the static SDW order. A CDW order seems to be absent in our sample. Upon cooling the system, a nodal gap in electronic excitations opens just when the strongest spin excitations start to diminish. It is therefore sufficient for the SDW fluctuations to slow down without completely freezing out in order to modify the band structure.

\section*{Methods}

For the Neutron scattering experiment, the sample was mounted on aluminum holder covered with Cd foils, and oriented in the (h,k,0) scattering plane. A Be filter was used to minimize contamination from high order monochromator Bragg reflections.  The scattered neutrons are recorded with a nine bladed graphite analyzer. All the blades are set to scatter neutrons at the same final energy of 5 meV, and direct the scattered neutrons through an adjustable radial collimator to different predefined areas on a position sensitive detector ~\cite{bahl2006inelastic,lefmann2006realizing}. This monochromatic q dispersive mode allows for an efficient mapping of magnetic excitations with an excellent q resolution.

Two types of scans were used: I) energy scan, in which the incoming neutrons energy is swept, and the $\textbf{q}$ information is embedded in the position of each blade. II) momentum scan, in which the incoming neutrons energy is fixed, the nine blades cover a small window in $\textbf{q}$, and the entire window is scanned. The contribution to a given $\textbf{q}$ is a weighted sum from the different blades.

Despite the Be filter, some contribution from the nuclear structure is unavoidable. For elastic scattering, this contribution survives to higher temperature than does the magnetic part, and therefore can be easily subtracted. For inelastic scattering, the contribution from phonons could not be subtracted, but it is expected to vary slowly with temperature close to the magnetic phase transitions. Therefore, all features in this scattering experiment which show abrupt temperature dependence around and below $T=50$~K are associated with the electronic (magnetic) system.


\section{Acknowledgments}

The Technion team is supported by the Israeli Science Foundation (ISF). G.D. is also funded by the Gordon and Betty Moore Foundation’s EPiQS Initiative through Grant GBMF4411. We thank the SINQ, Petra, and BESSY beam line staff for their excellent support. Parts of this research were carried out at the light source PETRA III at DESY, a member of the Helmholtz Association (HGF).

\end{document}